\documentclass[aps, prd, preprint, groupedaddress, nofootinbib]{revtex4-1}

\usepackage{graphicx}
\usepackage{epsfig}
\usepackage{bm}
\usepackage{amsfonts}
\usepackage{amsmath}
\usepackage{amssymb}
\usepackage{multirow}
\usepackage{subfigure}
\usepackage{wasysym}
\usepackage{array}
\usepackage{diagbox}
\usepackage{makecell}
\usepackage{slashbox}

\begin{document}



\title{Cosmological Implications of Modified Entropic Gravity}





\author{
Kazem Rezazadeh$^{2}$\footnote{kazem.rezazadeh@ipm.ir}
}

\affiliation{
\small{School of Astronomy, Institute for Research in Fundamental Sciences (IPM), P.O. Box 19395-5531, Tehran, Iran}
}


\date{\today}


\begin{abstract}

Taking into account the temperature corrections of the energy equipartition law for the bits of information that are coarse-grained on the holographic screen leads to a modification of Einstein's gravitational field equations. In the very high-temperature limit, which corresponds to strong gravitational fields, the modified gravitational equations reduce to the standard Einstein equations of general relativity, but in the low-temperature limit, which corresponds to the weak gravity regime, the modified equations show significant deviations from the standard Einstein equations. We solve the modified Einstein equations for the FRW metric and obtain the modified Friedmann equations. We see that the Friedmann equations obtained with this approach agree with the Friedmann equations previously obtained from the thermodynamic corrections of classical Newtonian mechanics. Using the modified Friedmann equations for a flat universe, we investigate the implications of our modified entropic cosmology (MEC) model. We show that our model can explain the dynamics of the universe without requiring any kind of dark energy. Using the Pantheon supernovae dataset, BAO data, Planck 2018 CMB data, and SH0ES measurements for $H_0$, we test the MEC model against observations. We will see that MEC fits the observational data better than the standard cosmological model of $\Lambda$CDM. We also see that our model can successfully solve the $H_0$ tension that challenges the standard cosmological model.

\end{abstract}



\maketitle



\section{Introduction}
\label{sec:introduction}

After the discovery of the accelerating expansion of the Universe in 1998 using data from type Ia supernovae \cite{SupernovaSearchTeam:1998fmf, SupernovaCosmologyProject:1998vns}, an important question arose as to what is the reason for this acceleration. The most prominent proposal to explain this acceleration is to include a dark energy component in the matter-energy content of the Universe, which, unlike conventional matter, has a negative pressure. In the standard model of cosmology, known as the $\Lambda$CDM model, the cosmological constant $\Lambda$ plays the role of dark energy which causes the current accelerating expansion of the Universe. The cosmological constant ($\Lambda$), first added to the equations of general relativity by Einstein, plausibly explains the current accelerating expansion. However, the cosmological constant is associated with significant theoretical and practical problems \cite{Weinberg:1988cp, Carroll:2000fy, Padmanabhan:2002ji}. One of the biggest problems is the ``fine-tuning problem of the cosmological constant,'' which refers to the huge discrepancy between the predicted value of this constant based on quantum theories and its observed value in the Universe. Theoretical calculations, such as the quantum vacuum energy, predict a much larger value, while cosmological observations show that the true value of the cosmological constant is much smaller. This huge discrepancy, described as ``the worst theoretical prediction in the history of physics,'' indicates that our current understanding of quantum physics and gravity is incomplete. Furthermore, the exact nature of dark energy and why it agrees with the observed value of the cosmological constant are still not fully understood, which is one of the greatest challenges facing modern cosmology.

Another problem related to the cosmological constant is the ``coincidence problem,'' \cite{Weinberg:1988cp, Peebles:2002gy} which implies: Why did dark energy begin to dominate at precisely the time in the history of the Universe that we can observe? In other words, the density of dark energy is now comparable to the density of matter in the Universe, while the two have varied differently throughout the history of the Universe. This strange synchronicity, sometimes described as a cosmic coincidence, raises the question of whether this is a coincidence or whether it reflects an unknown physical principle. These cosmological constant problems manifest some of the deepest mysteries in physics and cosmology.

Another problem that plagues the standard model of cosmology is the Hubble tension. This refers to the discrepancy between the two main methods of measuring the expansion rate of the Universe, known as the Hubble constant. This tension became apparent when the value of the Hubble constant was measured through local observations, such as type Ia supernovae and Cepheid variables \cite{Riess:2018byc, Riess:2019cxk, Riess:2021jrx, Riess:2022mme}, was inconsistent with the value obtained from observations of the cosmic microwave background (CMB) by the Planck satellite \cite{Planck:2018vyg, Planck:2019nip, Planck:2018lbu}. Local measurements, such as the baseline result of the SH0ES team from the Cepheid-SN sample estimate the Hubble constant to be $H_{0}=73.04\pm1.04\,\mathrm{km\ s^{-1}Mpc^{-1}}$ \cite{Riess:2021jrx}, while the Planck 2018 CMB data suggest a value of $H_{0}=67.4\pm0.5\,\mathrm{km\ s^{-1}Mpc^{-1}}$ \cite{Planck:2018vyg}. This discrepancy of about 8\%, which is statistically significant, is considered one of the greatest challenges facing modern cosmology. This tension may indicate a need to revise the standard cosmological model. However, the Hubble constant tension raises the question of whether there might be new physics beyond this model. Some of the possibilities raised include modifications in the properties of dark energy, the existence of unknown dark matter particles, or even modifications to general relativity. Resolving this tension would not only help us better understand the expansion rate of the Universe, but it could also open a window into the discovery of new and unknown physics. Various models have been proposed to resolve the Hubble tension or at least reduce it to some extent (see, e.g., \cite{Barreira:2014jha, Umilta:2015cta, DiValentino:2016hlg, DiValentino:2017rcr, Kumar:2016zpg, Poulin:2018cxd, Poulin:2021bjr, Davari:2019tni, Agrawal:2019lmo, Smith:2019ihp, Murgia:2020ryi, Karwal:2021vpk, Rezazadeh:2022lsf}).

The relationship between gravity and thermodynamics is one of the most fascinating and complex topics in theoretical physics, studied in the context of modern theories such as general relativity and quantum mechanics. One of the most important concepts in this field is the thermodynamics of black holes developed by physicists such as Hawking and Bekenstein \cite{Bardeen:1973gs, Hawking:1974rv, Hawking:1975vcx, Bekenstein:1973ur}. According to this theory, black holes have an entropy that is proportional to the area of their event horizon \cite{Hawking:1975vcx, Bekenstein:1973ur}. This idea suggests that gravity, as the dominant force on cosmic scales, interacts with the laws of thermodynamics. Furthermore, the Hawking radiation \cite{Hawking:1974rv, Hawking:1975vcx} emitted by black holes suggests a deeper connection between gravity and thermodynamics, as this phenomenon shows that black holes not only lose mass and energy but also generate entropy.

After Bekenstein and Hawking a lot of work has been done to disclose the connection between thermodynamics and gravity \cite{Davies:1974th, Unruh:1976db}. In 1995, Jacobson \cite{Jacobson:1995ab} showed that using the fundamental relation $\delta Q=TdS$ for the thermal energy passing through a local Rindler causal horizon, one can derive Einstein's gravitational equations as equations of state for space-time. Subsequent studies confirmed that there was a deep connection between the gravitational theories and the laws of thermodynamics \cite{Padmanabhan:2002sha, Padmanabhan:2009kr, Eling:2006aw, Akbar:2006mq}. In the context of cosmology, this connection allows one to derive Friedmann's first equation from the first law of thermodynamics for the apparent metric horizon FRW \cite{Akbar:2006kj, Cai:2005ra, Cai:2006rs, Wang:2001bf, Sheykhi:2007zp, Sheykhi:2007gi, Sheykhi:2008qs, Sheykhi:2009zv, Sheykhi:2010zz, Sheykhi:2013rla, Sheykhi:2015eul, Asghari:2021lzu, Teimoori:2023hpv}. This connection contributes to a better understanding of the nature of space-time and the fundamental laws of the universe.

Continuing to explore the connection between gravity and thermodynamics, we can mention the concept of the ``holographic principle'', which was first proposed by 't Hooft \cite{tHooft:1993dmi} and then developed by Susskind and others \cite{Susskind:1994vu, Susskind:2005js, Witten:2001kn, Padmanabhan:2006fn, Sheikhahmadi:2021jkz} in cosmology. This principle states that information about a volume of space-time can be encoded on the boundary of that volume (such as the event horizon of a black hole). The idea originated from studies of the thermodynamics of black holes and suggests that gravity and thermodynamics may be two sides of the same coin. In other words, the laws of thermodynamics that usually apply to ordinary physical systems can also apply on cosmic scales. This deeper connection is also explored in theories of quantum gravity, such as string theory and loop quantum gravity, and suggests that thermodynamics may be the key to understanding the fundamental nature of gravity and the structure of space-time. These ideas not only deepen our understanding of black holes but also help solve some of the greatest puzzles in modern physics, such as the black hole information paradox \cite{Hawking:1976ra, Preskill:1992tc}.

Another notable achievement in the connection between gravity and thermodynamics is the conceptual theory of the entropic force, developed by Verlinde \cite{Verlinde:2010hp}, which proposes that gravity is no longer a fundamental force, but an emergent force. According to Verlinde, one can start from first principles and gravity naturally emerges as a consequence of entropic force. This entropic force originates from changes in the amount of information about the position of material bodies. In Verlinde's approach, the holographic principle and the equipartition law of energy play important roles. By studying the statistical mechanics of bits of information stored on the boundary of a thermodynamic system, Verlinde was able to derive Newton's second law as well as Newton's law of universal gravitation. By extending his approach to the relativistic case, he also derived Einstein's equations of general relativity using the concept of the entropic force \cite{Verlinde:2010hp}.

Many studies have been conducted so far on gravity as an entropic force and its cosmological consequences (see, e.g., \cite{Cai:2010hk, Cai:2010sz, Smolin:2010kk, Li:2010cj, Tian:2010uy, Modesto:2010rm, Sheykhi:2010wm, Hendi:2010xr, Kiselev:2010mz, Myung:2010jv, Banerjee:2010yd, Gao:2010fw, Wei:2010am, Chang:2010be, Easson:2010av, Easson:2010xf, Zhang:2011uf, Mann:2011rh, Sheykhi:2012vf, Sheykhi:2011sqb, Garcia-Bellido:2021idr, Arjona:2021uxs, Chagoya:2023hjw, Gohar:2023hnb, Gohar:2023lta, Kibaroglu:2025hzc}). In Ref. \cite{Sheykhi:2012vf}, it has been shown that by considering the Debye correction to the equipartition law of energy, a modified form of Newton's law of gravity, modified Einstein's equations, and modified Poisson's equation can be derived. In particular, in this and other works, it has been shown that considering the Debye correction in the low-temperature limit, which is related to weak gravitational fields, can provide a logical justification for the basis of MOND theory. In Ref. \cite{Sheykhi:2011sqb}, the authors have derived the modified Friedmann equations by considering the Debye temperature correction. For this purpose, they have used a thermodynamic approach based on temperature corrections to the laws of Newtonian classical mechanics. However, it would be very interesting to derive the Friedmann equations in a general relativistic approach by applying the modified Einstein field equations for the FRW metric, which is one of the main goals of this paper.

In this paper, we intend to present a cosmological model based on the connection between gravity and thermodynamics, with which we can answer some of the problems of the standard model of cosmology. For this purpose, we consider temperature corrections to the equipartition law of energy. Studies of the statistical mechanics of condensed matters have revealed that it is necessary to consider corrections to the equipartition law of energy to justify their heat capacity based on the statistical behavior of bosons at low temperatures \cite{Reif1965, Pathria:1996hda}. This correction is known in the study of solid-state physics as the Debye correction, which is based on considering certain assumptions for the period of oscillation of phonons and well justifies the temperature dependencies of the heat capacity of solids at low temperatures \cite{Reif1965, Pathria:1996hda}.

These achievements motivate us to consider temperature corrections to the equipartition law of energy which are applied to the bits of information coarse-grained on the holographic screen. The inclusion of these corrections leads to the modification of Einstein's gravitational field equations. In the high-temperature limit, which corresponds to strong gravitational fields, the modified equations reduce to Einstein's standard equations of general relativity. However, in the low-temperature limit, which corresponds to very weak gravitational fields, the modified gravitational equations show a substantial deviation from Einstein's standard gravity.

In the next step, we solve the modified Einstein equations for the FRW metric and extract the modified Friedmann equations. We show that the results of our general relativistic approach are in agreement with the findings based on the thermodynamical corrections to the laws of Newtonian mechanics \cite{Sheykhi:2011sqb}. We then apply the modified Friedmann equations for a flat Universe to scrutinize the cosmological consequences of our modified gravity theory. We demonstrate that the modified entropic cosmology (MEC) model can explain the current cosmic acceleration without demanding any kind of dark energy component. We then constrain our cosmological model using observational data from different sources. In our work, we include data from the Pantheon sample of supernovae Ia \cite{Pan-STARRS1:2017jku}. We also include the BAO data from the measurements of \cite{Beutler:2011hx, SDSS:2009ocz, Blake:2011wn}. We further include the latest CMB data from the Planck group \cite{Planck:2018vyg, Planck:2019nip, Planck:2018lbu}. Additionally, we will include the measurement of $H_0$ provided by the Supernovae and H0 for the Equation of State (SH0ES) project \cite{Riess:2021jrx}.

The remainder of this paper is structured as follows. In Sec. \ref{sec:setup}, we introduce our modified entropic gravity model and present the modified Einstein equations in this scenario. Then, in this section, we derive the modified Friedmann equations from the modified Einstein equations in our setup. In Sec. \ref{sec:data}, we introduce the datasets that are implemented in our investigation. Then, in Sec. \ref{sec:results}, we present the results of our MCMC analysis and discuss our findings. Finally, in Sec. \ref{sec:conclusions}, we summarize our conclusions.


\section{Modified Entropy Gravity}
\label{sec:setup}

Here, we first review briefly the approach of \cite{Sheykhi:2012vf}, for derivation of modified Einstein equations with the temperature corrections. For this purpose, we consider a thermodynamic system with three spatial dimensions, separated from the environment by a two-dimensional boundary. The holographic principle states that information about the position of material bodies inside this system is stored on its two-dimensional boundary. The number of bits of information on the boundary, which acts like a holographic screen, is given by \cite{Verlinde:2010hp}
\begin{equation}
\label{N}
N=\frac{A}{\ell_{P}^{2}} \, ,
\end{equation}
where $A$ is the area of the two-dimensional boundary and $\ell_{P}=\sqrt{\hbar G/c^{3}}$ is the Planck length. In our notation, the Planck mass is defined as $m_{P}=\sqrt{\hbar c/G}$. In addition, the reduced Planck mass is defined as $M_{P}\equiv m_{P}/(8\pi)$. Throughout this paper, we work in dimensions where $c=\hbar=k_{B}=1$, but it is convenient to still keep the explicit symbol of these quantities in some equations.

The holographic screen in our setup behaves like a casual horizon and has the following entropy
\begin{equation}
\label{S}
S=\frac{N}{4}=\frac{A}{4\ell_{P}^{2}} \, .
\end{equation}
The number of degrees of freedom of the system is equal to the number of bits of information on the holographic screen. Therefore, the equipartition law of energy for these degrees of freedom, by taking into account the temperature corrections, can be written as follows
\begin{equation}
\label{E}
E=\frac{1}{2}Nk_{B}Tf(T) \, .
\end{equation}
In this relation, $T$ represents the horizon temperature, and $f(T)$ is a general function of temperature that includes the temperature corrections in the energy equipartition law relation. In the case of $f(T) = 1$, Eq  \eqref{E} reduces to the standard equipartition law. The idea of considering the temperature correction in the equipartition law of energy came from the studies of the bosonic statistical mechanics of phonons in crystals in the context of solid state physics \cite{Reif1965, Pathria:1996hda}. Studies in solid state physics reflect the fact that by considering the Debye temperature correction in Eq  \eqref{E}, one can successfully describe the behavior of the heat capacity of crystals at low temperatures \cite{Reif1965, Pathria:1996hda}. In this study, we consider a general function in relation \eqref{E} which is generally different from the Debye function.

The temperature of the holographic screen is given by the Unruh temperature formula, which states that an observer in an accelerating frame experiences the temperature of \cite{Unruh:1976db}
\begin{equation}
\label{T}
T=\frac{\hbar\alpha}{2\pi k_{B}c} \, ,
\end{equation}
where $\alpha$ represents the magnitude of the acceleration. In general relativity, the acceleration vector is related to the Newtonian potential as follows
\begin{equation}
\label{alpha}
\alpha^{\mu}=-\nabla^{\mu}\phi \, .
\end{equation}
The quantity $\phi$ can be considered as a generalization of the Newtonian potential in general relativity. The holographic screen is considered as a closed surface $\mathcal{S}$ which is an equipotential surface with the same $\phi$ at all its points. The Newtonian potential $\phi$ is related to the Killing vector as follows \cite{Wald:1984rg}
\begin{equation}
\label{phi}
\phi=\frac{1}{2}\ln\left(-\xi^{\mu}\text{\ensuremath{\xi}}_{\mu}\right) \, .
\end{equation}
Here $\xi^{\mu}$ is the general timelike Killing vector that fulfills in the Killing equation \cite{Wald:1984rg},
\begin{equation}
\label{Killing}
\nabla_{\mu}\xi_{\nu}+\nabla_{\nu}\xi_{\mu}=0 \, .
\end{equation}
Also, the Killing vector is related to the Ricci tensor $R_{\mu}^{\nu}$ via the following relation \cite{Wald:1984rg}
\begin{equation}
\label{xinu-Rmunu}
\nabla^{\mu}\nabla_{\mu}\xi^{\nu}=-R_{\mu}^{\nu}\xi^{\mu} \, .
\end{equation}

Following the approach of \cite{Sheykhi:2012vf}, we can obtain the following relation for the total mass enclosed by the holographic screen
\begin{equation}
\label{M-xi}
M=-\frac{1}{8\pi G}\oint_{\mathcal{S}}\nabla^{\mu}\xi^{\nu}f(T)d\mathcal{S}_{\mu\nu} \, .
\end{equation}
Using Stokes' theorem, we can convert the integral over the enclosed surface to the integral over the enclosed volume,
\begin{equation}
\label{M-Rmunu}
M=\frac{1}{4\pi G}\int_{\Sigma}\left(f(T)R_{\mu\nu}-e^{-2\phi}\xi_{\nu}\nabla_{\mu}\xi^{\lambda}\nabla_{\lambda}f(T)\right)n^{\mu}\xi^{\nu}d\Sigma \, .
\end{equation}
where $\Sigma$ indicates the whole volume enclosed by the closed surface $\mathcal{S}$. Also, $n^\mu$ is a unit vector which is normal to the hypersurface $\Sigma$.

On the other hand, the mass $M$ can be expressed as an integral of certain components of stress-energy tensor $\mathcal{T}_{\mu\nu}$ over the volume $\Sigma$ \cite{Wald:1984rg},
\begin{equation}
\label{M-Tmunu}
M=2\int\left(\mathcal{T}_{\mu\nu}-\frac{1}{2}\mathcal{T}g_{\mu\nu}\right)n^{\mu}\xi^{\nu}d\Sigma  \, ,
\end{equation}
where $\mathcal{T}$ is the trace of the stress-energy tensor.

By equating the two equations \eqref{M-Rmunu} and \eqref{M-Tmunu}, the modified Einstein equations in our entropic gravity scenario are obtained as follows \cite{Sheykhi:2012vf}
\begin{equation}
\label{Einstein}
f(T)R_{\mu\nu}-e^{-2\phi}\text{\ensuremath{\xi}}_{\nu}\nabla_{\mu}\xi^{\lambda}\partial_{\lambda}f(T)=-8\pi G\left(\mathcal{T}_{\mu\nu}-\frac{1}{2}\mathcal{T}g_{\mu\nu}\right) \, .
\end{equation}
In this equation, we see that considering the temperature corrections causes the geometry part of the Einstein equations to be modified while the matter-energy content part which is the source of gravitation is left untouched. In the limit of strong gravitational fields, we have $f(T) = 1$, and the above equations become the standard Einstein equations. But in the regime of very weak gravitational fields, the effects of temperature corrections may cause the gravitational theory to be completely different from Einstein's general relativity.

Next, we want to obtain the solution of the modified Einstein equations \eqref{Einstein} for the homogeneous and isotropic FRW metric with the line element
\begin{equation}
\label{FRW}
ds^{2}\equiv g_{\mu\nu}dx^{\mu}dx^{\nu}=dt^{2}-a^{2}(t)\left[\frac{dr^{2}}{1-kr^{2}}+r^{2}d\theta^{2}+r^{2}\sin^{2}(\theta)\right] \, .
\end{equation}
In this relation, $k$ denotes the curvature constant and characterizes the spatial curvature of space-time. Also, $a(t)$ is the scale factor of the Universe used to express the size of the Universe at any time. The Hubble parameter is defined as $H \equiv \dot{a}/a$ and characterizes the expansion rate of the Universe.

For the homogeneous and isotropic FRW metric, the acceleration depends only on the time coordinate, so the Unruh temperature \eqref{T} becomes $T=\ddot{a}/2\pi$, where $\ddot{a}$ is equal to the acceleration experienced by the observer located in the vicinity of the holographic screen. Therefore, we can take the function which is used for the temperature correction of the equipartition law \eqref{E} to be a function of acceleration, $f = f(\ddot{a})$.

The components of the Ricci tensor for the metric \eqref{FRW} are given by
\begin{align}
\label{R00}
R_{00} &= 3\frac{\ddot{a}}{a} \, ,
\\
\label{R11}
R_{11} &= -\frac{2k+a\ddot{a}+2\dot{a}^{2}}{1-kr^{2}} \, ,
\\
\label{R22}
R_{22} &= -r^{2}\left(a\ddot{a}+2\dot{a}^{2}+2k\right) \, ,
\\
\label{R33}
R_{33} &= -r^{2}\sin^{2}(\theta)\left(a\ddot{a}+2\dot{a}^{2}+2k\right) \, .
\end{align}
Using these equations, the Ricci scalar will be as follows
\begin{equation}
\label{R}
R\equiv R_{\mu}^{\mu}=\frac{6}{a^{2}}\left(a\ddot{a}+\dot{R}^{2}+k\right) \, .
\end{equation}

We consider the matter-energy content of the Universe as a perfect fluid with the following stress-energy tensor
\begin{equation}
\label{Tmunu}
\mathcal{T}_{\nu}^{\mu}=\mathrm{diag}\left(\rho,-p,-p,-p\right) \, .
\end{equation}
where $\rho$ and $p$ denote the energy density a pressure of the fluid, respectively.

Now, we should determine the components of the Killing vector $\xi^{\mu}$. For this purpose, it is necessary to solve the Killing equation \eqref{Killing}. For the general Killing vector $\xi^{\mu}=\left(\xi^{0},\xi^{1},\xi^{2},\xi^{3}\right)$, the different components of the Killing equation \eqref{Killing} for the FRW metric \eqref{FRW} lead to the following equations
\begin{align}
\label{eqtt}
\mathrm{eq}_{tt} &:= \partial_{t}\xi^{0}=0 \, ,
\\
\label{eqtr}
\mathrm{eq}_{tr} &= \mathrm{eq}_{rt}:=\frac{a^{2}\partial_{t}\xi^{1}+\partial_{r}\xi^{0}\left(1-kr^{2}\right)}{1-kr^{2}}=0 \, ,
\\
\label{eqttheta}
\mathrm{eq}_{t\theta} &= \mathrm{eq}_{\theta t}:=a^{2}r^{2}\partial_{t}\xi^{2}+\partial_{\theta}\xi^{0}=0 \, ,
\\
\label{eqtphi}
\mathrm{eq}_{t\phi} &= \mathrm{eq}_{\phi t}:=a^{2}r^{2}\sin^{2}(\theta)\partial_{t}\xi^{3}+\partial_{\phi}\xi^{0}=0 \, ,
\\
\label{eqrr}
\mathrm{eq}_{rr} &:= \frac{2a}{1-kr^{2}}\left(-a\partial_{r}\xi^{1}+akr^{2}\partial_{r}\xi^{1}-akr\xi^{1}-\dot{a}\xi^{0}+\dot{a}kr^{2}\xi^{0}\right)=0 \, ,
\\
\label{eqrtheta}
\mathrm{eq}_{r\theta} &= \mathrm{eq}_{\theta r}:=\frac{a^{2}}{1-kr^{2}}\left(r^{2}\partial_{r}\xi^{2}-kr^{4}\partial_{r}\xi^{2}+\partial_{\theta}\xi^{1}\right)=0 \, ,
\\
\label{eqrphi}
\mathrm{eq}_{r\phi} &= \mathrm{eq}_{\phi r}:=\frac{a^{2}}{1-kr^{2}}\left(r^{2}\sin^{2}(\theta)\partial_{r}\xi^{3}-kr^{4}\sin^{2}(\theta)\partial_{r}\xi^{3}+\partial_{\phi}\xi^{1}\right)=0 \, ,
\\
\label{eqthetatheta}
\mathrm{eq}_{\theta\theta} &:= 2a^{2}r^{2}\partial_{\theta}\xi^{2}+2a\dot{a}r^{2}\xi^{0}+2a^{2}r\xi^{1}=0 \, ,
\\
\label{eqthetaphi}
\mathrm{eq}_{\theta\phi} &= \mathrm{eq}_{\phi\theta}:=a^{2}r^{2}\sin^{2}(\theta)\partial_{\theta}\xi^{3}+a^{2}r^{2}\partial_{\phi}\xi^{2}=0 \, ,
\\
\label{eqphiphi}
\mathrm{eq}_{\phi\phi} &:= 2a^{2}r^{2}\sin^{2}(\theta)\partial_{\phi}\xi^{3}+2a\dot{a}r^{2}\sin^{2}(\theta)\xi^{0}+2a^{2}r\sin(\theta)\left(\sin(\theta)\xi^{1}+r\cos(\theta)\xi^{2}\right)=0 \, .
\end{align}
By solving the above set of equations, the components of the Killing vector for the FRW metric are determined as follows \cite{HariDass:2015plk}
\begin{align}
\label{xi0}
\xi^{0} &= 0 \, ,
\\
\label{xi1}
\xi^{1} &= \sqrt{1-kr^{2}}\left[\sin(\theta)\left(C_{1}\cos(\phi)+C_{2}\sin(\phi)\right)+C_{3}\cos(\theta)\right] \, ,
\\
\label{xi2}
\xi^{2} &= \frac{\sqrt{1-kr^{2}}}{r}\left[\cos(\theta)\left(C_{1}\cos(\phi)+C_{2}\sin(\phi)\right)-C_{3}\sin(\theta)\right]+\left(C_{4}\sin(\phi)-C_{5}\cos(\phi)\right) \, ,
\\
\label{xi3}
\xi^{3} &= \frac{\sqrt{1-kr^{2}}}{r}\frac{1}{\sin(\theta)}\left(C_{2}\cos(\phi)-C_{1}\sin(\phi)\right)+\cot(\theta)\left(C_{4}\cos(\phi)+C_{5}\sin(\phi)\right)-C_{6} \, ,
\end{align}
where $C_1$-$C_6$ are constants.

Using the Killing vector obtained above in the component ``$tt$'' of the modified Einstein equation \eqref{Einstein}, we find
\begin{equation}
\label{Friedmann2}
\frac{\ddot{a}}{a}=-\frac{4\pi G}{3f(\ddot{a})}\left(\rho+3p\right) \, .
\end{equation}
This is the second Friedmann equation in the setup of modified entropic cosmology (MEC). This equation is in agreement with the result of \cite{Sheykhi:2011sqb}, in which the modified Friedmann equations were derived from considering the temperature corrections in Newton's laws. It is important to note that in our result the general function $f(\ddot{a})$ replaces the Debye function $\mathcal{D}(x)$, which was used as the temperature correction function in the analysis of \cite{Sheykhi:2011sqb}.

From the component ``$rr$'' of the modified Einstein equation \eqref{Einstein}, we arrive at the following result
\begin{equation}
\label{Einstein_rr}
f(\ddot{a})\left(a\ddot{a}+2\dot{a}^{2}+2k\right)+4\pi Ga^{2}\left(p-\rho\right)=0 \, .
\end{equation}
By combining this equation with Eq. \eqref{Friedmann2}, we acquire the first Friedmann equation in the MEC model as
\begin{equation}
\label{Friedmann1}
H^{2}+\frac{k}{a^{2}}=\frac{8\pi G}{3f(\ddot{a})}\rho \, .
\end{equation}
This result is consistent with the equation derived in \cite{Sheykhi:2011sqb} but with a little deference. In the result of \cite{Sheykhi:2011sqb} for the modified first Friedmann equation, on the right-hand side of the equation, the inverse of the temperature correction function appears under an integral, and if this function is constant or slow-varying with respect to the integration parameter, then their result turns into our result. This difference arises from that in our general relativistic approach no integration is required to extract the Friedmann equations, but in the approach of \cite{Sheykhi:2011sqb}, it is necessary to take the integral from the second Friedmann equation, while the continuity equation is also used.

Now that we have derived the modified Friedmann equations in our entropic gravity scenario, we can examine the cosmological consequences of this scenario. For this purpose, we will continue to consider the space curvature of the Universe as flat ($k = 0$). We also assume that the matter-energy of the Universe is comprised of baryon, cold dark matter, and radiation. Thus, the energy density of the Universe is given by
\begin{equation}
\label{rho}
\rho=\rho_{b}+\rho_{c}+\rho_{r} \, .
\end{equation}
In this relation, $\rho_{b}$, $\rho_{c}$, and $\rho_{r}$ are the energy densities of baryons, cold dark matter, and radiation, respectively. Note that this equation does not include the cosmological constant or any other dark energy contribution. Since the pressure of the baryon and dark matter components vanishes, therefore only the radiation pressure contribution remains in the total pressure relation,
\begin{equation}
\label{p}
p = p_{r} = \frac{1}{3}\rho_{r} \, .
\end{equation}

From the continuity equations for the individual components of the Universe, the variations of the energy densities with a scale factor are obtained as follows
\begin{align}
\label{rhob}
\rho_{b} &= \rho_{b0}a^{-3} \, ,
\\
\label{rhoc}
\rho_{c} &= \rho_{c0}a^{-3} \, ,
\\
\label{rhor}
\rho_{r} &= \rho_{r0}a^{-4} \, .
\end{align}
For each component ``$i$'' of matter-energy content, the normalized density parameter in our scenario is defined as
\begin{equation}
\label{Omegati0}
\tilde{\Omega}_{i0}\equiv\rho_{i0}/3M_{P}^{2}f_{0}H_{0}^{2} \, .
\end{equation}
The sum of the normalized energy densities is unity,
\begin{equation}
\label{sum_Omegati0}
\underset{i}{\Sigma}\tilde{\Omega}_{i0}=1 \, .
\end{equation}

For comparing the results of our cosmology model with the observational data, it is much better to use the conventional density parameter instead of the normalized density parameter. The conventional density parameter is defined as
\begin{equation}
\label{Omegai0}
\Omega_{i0}\equiv\frac{\rho_{i0}}{3M_{P}^{2}H_{0}^{2}} \, .
\end{equation}
It is important to note that, in contrast to the sum of the normalized density parameters, the sum of the conventional density parameters in our scenario is opposite to unity and instead is equal to
\begin{equation}
\label{sum_Omegai0}
\underset{i}{\Sigma}\Omega_{i0}=f_{0} \, ,
\end{equation}
where $f_0$ is the value of the function of temperature correction at the present time.

In our study, we consider neutrinos to be massless. Therefore, the radiation component here consists of two components: photons and massless neutrinos. For the energy density of photons, we have \cite{WMAP:2012nax}
\begin{equation}
\label{Omegagamma0}
\Omega_{\gamma0}=2.469\times10^{-5}h^{-2} \, .
\end{equation}
Including the contribution of massless neutrinos, the radiation energy density will be
\begin{equation}
\label{Omegar0}
\Omega_{r0}=\left(1+0.2271N_{\mathrm{eff}}\right)\Omega_{\gamma0} \, ,
\end{equation}
where $N_{\mathrm{eff}}$ is the effective number of neutrinos, and we take it to be $N_{\mathrm{eff}}=3.046$ according to the Standard Model of particle physics \cite{Mangano:2005cc}.

In the following, we define the parameter $\alpha$ as the absolute magnitude of the acceleration, $\alpha\equiv\left|\ddot{a}\right|$. Since this quantity has dimensions, it is more appropriate to normalize it as $\tilde{\alpha}\equiv\left|\ddot{a}\right|/H_{0}^{2}$. This parameter is dimensionless and so much easier to work with. In addition, we define the variable $x\equiv\tilde{\alpha}/\tilde{\alpha}_{c}$. In this relation, $\tilde{\alpha}_{c}=\tilde{\alpha}(a_{c})$ is the normalized acceleration at the critical scale factor $a_{c}$. The critical normalized acceleration $\tilde{\alpha}_{c}$ determines the epoch at which the transition from standard Einstein gravity to modified entropic gravity occurs. With these definitions, we can now specify the temperature correction function which appears in the energy equipartition law \eqref{E}. We suppose this function is an exponential function in the form of
\begin{equation}
\label{f}
f(x)\equiv\exp\left(-x^{-n}\right) \, ,
\end{equation}
where $n$ is a positive constant parameter. In the limit of strong gravitational fields, the magnitude of the normalized acceleration is much larger than the critical normalized acceleration ($\tilde{\alpha}\gg\tilde{\alpha}_{c}$), and as a result we have $x\to\infty$. In this limit, we have $f(x)\approx1$, so, the effective theory of gravity reduces to Einstein's standard gravity. On the other hand, in the limit of very weak gravitational fields, we have $\tilde{\alpha}\ll\tilde{\alpha}_{c}$, and therefore $x\to0$. In this limit, we have $f(x)\to0$. Thus, in the limit of very weak gravitational fields, the theory of modified entropic gravity dominates, and hence significant deviations from Einstein's general relativity are expected.

By substituting the function $f(x)$ from equation \eqref{f} into the second Friedmann equation \eqref{Friedmann2}, we arrive at the following equation
\begin{equation}
\label{alphat-eq}
\tilde{\alpha}\exp\left[\left(\frac{\tilde{\alpha}}{\tilde{\alpha}_{c}}\right)^{n}\right]=\frac{1}{2a^{3}}\left[\left(\text{\ensuremath{\Omega}}_{b0}+\text{\ensuremath{\Omega}}_{c0}\right)a+2\text{\ensuremath{\Omega}}_{r0}\right] \, .
\end{equation}
The solution to this equation is as follows
\begin{equation}
\label{alphat}
\tilde{\alpha}=n^{1/n}\tilde{\alpha}_{c}W^{-1/n}\left[2^{n}na^{-3n}\tilde{\alpha}_{c}^{n}\left(\left(\text{\ensuremath{\Omega}}_{b0}+\text{\ensuremath{\Omega}}_{c0}\right)a+2\text{\ensuremath{\Omega}}_{r0}\right)^{-n}\right] \, ,
\end{equation}
where we have used the Lambert $W$ function defined as solution of the equation $We^{W}=x$. By substituting this solution into the first Friedmann equation \eqref{Friedmann1}, we reach
\begin{equation}
\label{Ht}
\tilde{H}=\exp\left[\frac{1}{2}\left(\frac{\tilde{\alpha}_{c}}{\tilde{\alpha}}\right)^{n}\right]\sqrt{\left(\text{\ensuremath{\Omega}}_{b0}+\text{\ensuremath{\Omega}}_{c0}\right)a^{-3}+\text{\ensuremath{\text{\ensuremath{\Omega}}_{r0}}}a^{-4}} \, ,
\end{equation}
where $\tilde{H}\equiv H/H_{0}$ is the is normalized Hubble parameter.

To determine the parameter $\tilde{\alpha}_{c}$, we use the condition that the present-day value of the normalized Hubble parameter for a flat Universe must be equal to unity, $\tilde{H}(a=1)=1$. By imposing this equation, the validity of condition \eqref{sum_Omegai0} is guaranteed for the sum of the conventional density parameters in our model. To solve this equation for each given set of parameters, we need to use a numerical root-finding. After determining the value of $\tilde{\alpha}_{c}$, we utilize another numerical root-finding to solve the equation $\tilde{\alpha}(a_{c})=\tilde{\alpha}_{c}$. In this way, we determine the value of $a_{c}$ for each set of free parameters of the model. Thus, as is clear, the parameters $\tilde{\alpha}_{c}$ and $a_{c}$ in our analysis are not free parameters, but rather derived parameters.


\section{Observational Data}
\label{sec:data}

In this section, we evaluate the validity of our cosmology model in comparison to the observational data. In our work, we use SN, BAO, and CMB data together with the SH0ES measurement for $H_0$. In the following subsections, we discuss each of these datasets briefly. More details about these datasets and their analysis can be found in \cite{Nesseris:2012tt, Mehrabi:2015hva, Rabiei:2015pha, Harko:2022unn, Pooya:2023bwh, Shahhoseini:2025sgl, Pourojaghi:2025vmj} and references therein.


\subsection{SN Data}
\label{subsec:SN}

Type Ia supernovae are one of the most important research topics in the study of cosmic background dynamics due to their standard nature, and they still provide significant constraints on late-time cosmic evolution. The SN dataset is simply the difference between the apparent and absolute magnitudes of observed supernovae at redshift $z$, called the distance modulus, which is theoretically given by the equation
\begin{equation}
\label{muth}
\mu_{\mathrm{th}}(z)=5\log_{10}d_{L}(z)-\mu_{0} \, ,
\end{equation}
where $\mu_{0}=42.384-5\log_{10}h$. Here, $d_{L}$ is the luminosity distance, defined as
\begin{equation}
\label{dL}
d_{L}(z)=(1+z)\int_{0}^{z}\frac{dz'}{H(z')} \, .
\end{equation}

In this work, we use the SN Ia dataset from the Pantheon sample \cite{Pan-STARRS1:2017jku} which consists of 1048 data points. The value of $\chi^2$ for this dataset is calculated from the following equation
\begin{equation}
\label{chi2SN}
\chi_{\mathrm{SN}}^{2}=\stackrel[i=1]{n}{\sum}\frac{\left[\mu_{\mathrm{th}}\left(z_{i}\right)-\mu_{\mathrm{obs}}\left(z_{i}\right)\right]^{2}}{\sigma_{\mu,i}^{2}} \, .
\end{equation}
where $\mu_{\mathrm{th}}\left(z_{i}\right)$ is the theoretical prediction of the distance modulus at redshift $z_i$, and $\mu_{\mathrm{obs}}\left(z_{i}\right)$ is the distance modulus measured from observations. Furthermore, $\sigma_{\mu,i}$ represents the uncertainty in measurement of each data point.


\subsection{BAO Data}
\label{subsec:BAO}

In recent years, research on baryon acoustic oscillations has revealed that these observations can provide a useful geometric probe for dark energy or modified gravity models. The position of the BAO peaks in the CMB power spectrum depends on the ratio of the angular diameter distance $D_{v}(z)$ to the size of the sound horizon $r_s(z)$ at the drag redshift $z_d$, at which the baryons are released from the photons. Komatsu et al. \cite{WMAP:2008lyn} have shown that the drag epoch takes place slightly after the decoupling epoch, $z_d < z_*$. Since the baryons are affected by the potential well, the size of the sound horizon during the drag period is slightly larger than its size during the decoupling period.

The acoustic horizon is calculated with the following equation
\begin{equation}
\label{rs}
r_{s}\left(z_{d}\right)=\frac{c}{H_{0}}\int_{z_{d}}^{\infty}\frac{dz}{\tilde{H}(z)\sqrt{3\left[1+\frac{3\Omega_{b0}}{4(1+z)\Omega_{\gamma0}}\right]}} \, ,
\end{equation}
where $\tilde{H}$ is the Hubble parameter of the model and its relation for the modified entropic cosmology (MEC) is given in Eq. \eqref{Ht}. For the angular diameter distance, we use the following equation \cite{SDSS:2009ocz, Nesseris:2012tt}
\begin{equation}
\label{Dv}
D_{v}(z)=\left[(1+z)^{2}D_{A}^{2}(z)\frac{cz}{H(z)}\right]^{1/3} \, .
\end{equation}
To compare with the observations, we should calculate the following ratio in our theoretical model
\begin{equation}
\label{dz}
d_{z}(z)=\frac{r_{s}\left(z_{d}\right)}{D_{v}\left(z_{\mathrm{eff}}\right)} \, .
\end{equation}

In this study, we use the BAO dataset from 6dF \cite{Beutler:2011hx}, SDSS \cite{SDSS:2009ocz}, and WiggleZ \cite{Blake:2011wn} measurements. These data are presented in Table \ref{table:BAO}. The $\chi^2$ for this dataset is computed by using
\begin{equation}
\label{chi2BAO}
\chi_{BAO}^{2}=\underset{i,j}{\sum}\left(\left.d_{z}\left(z_{i}\right)\right|_{\mathrm{th}}-\left.d_{z}\left(z_{i}\right)\right|_{\mathrm{obs}}\right)C_{ij}^{-1}\left(\left.d_{z}\left(z_{j}\right)\right|_{\mathrm{th}}-\left.d_{z}\left(z_{j}\right)\right|_{\mathrm{obs}}\right) \, .
\end{equation}
In this equation, $C_{ij}^{-1}$ is the inverse of the covariance matrix and it is given by \cite{Blake:2011wn}
\begin{equation}
\label{invcovBAO}
C_{ij}^{-1}=\left(\begin{array}{cccccc}
4444 & 0 & 0 & 0 & 0 & 0\\
0 & 30318 & -17312 & 0 & 0 & 0\\
0 & -17312 & 87046 & 0 & 0 & 0\\
0 & 0 & 0 & 23857 & -22747 & 10586\\
0 & 0 & 0 & -22747 & 128729 & -59907\\
0 & 0 & 0 & 10586 & -59907 & 125536
\end{array}\right) \, .
\end{equation}

\begin{table*}[!ht]
\centering
\caption{The BAO data \cite{Beutler:2011hx, SDSS:2009ocz, Blake:2011wn} which are included in our analysis. The inverse covariance Matrix of these data is given by Eq. \eqref{invcovBAO}.}
\scalebox{1.0}{
\begin{tabular}{|l|c|c|}
\hline
\hline
Survey $\qquad$ & $\qquad$ $z_{\mathrm{eff}}$ $\qquad$ & $\qquad$ $d_{z}$ $\qquad$\tabularnewline
\hline
6dF & $0.106$ & $0.336\pm0.015$\tabularnewline
SDSS & $0.2$ & $0.1905\pm0.0061$\tabularnewline
SDSS & $0.35$ & $0.1097\pm0.0036$\tabularnewline
WiggleZ & $0.44$ & $0.0916\pm0.0071$\tabularnewline
WiggleZ & $0.6$ & $0.0726\pm0.0034$\tabularnewline
WiggleZ & $0.73$ & $0.0592\pm0.0032$\tabularnewline
\hline
\end{tabular}
}
\label{table:BAO}
\end{table*}


\subsection{CMB Data}
\label{subsec:CMB}

Using CMB temperature and polarization data to constrain a cosmological model requires that the modified Boltzmann equations in that model be derived and fully solved. Deriving the Boltzmann equations in turn requires that the cosmological perturbation theory for the model in question be examined. Since the gravitational theory in our model is completely different from the standard Einstein theory of gravity throughout the evolution of the Universe, the Boltzmann equations in our model will be quite different from the standard Boltzmann equations used in common cosmological codes such as CAMB \cite{Lewis:1999bs} and CLASS \cite{Blas:2011rf}. Therefore, to be able to use these codes in our work, we need to obtain the modified Boltzmann equations within the framework of our model. Doing this requires examining the cosmological perturbation theory completely for the modified Einstein equations of Eq. \eqref{Einstein}. As a result, it is necessary to modify many parts of the Boltzmann codes. This is a heavy undertaking and is beyond the scope of the current research and will be left to future research. Alternatively, one could use reduced CMB data that only depends on the dynamics of the cosmological background. The details of the extraction of CMB data for the 2018 Planck data \cite{Planck:2018vyg, Planck:2019nip, Planck:2018lbu} are described in Ref. \cite{Chen:2018dbv} and we summarize the results below.

The use of the position of the CMB acoustic peak is useful for constraining dark energy models or modified gravity models because this position depends on the dynamics of the Universe through the angular diameter distance. The position of this peak in the CMB thermal anisotropy power spectrum is specified by the quantities $\{R,\,l_{a},\,\Omega_{b0}h^{2}\}$. Here, $R(z_{*})$ is the scale distance or redshift parameter at the epoch of decoupling, which is given by the following equation
\begin{equation}
\label{Rzstar}
R(z_{*})=\frac{1}{c}\sqrt{\Omega_{m0}}H_{0}\left(1+z_{*}\right)D_{A}(z_{*}) \, .
\end{equation}
The parameter $z_{*}$ represents the decoupling redshift, and we use the fitting formula given in Ref. \cite{Hu:1995en} for it. Also, $D_{A}(z_{*})$ is the angular diameter distance,
\begin{equation}
\label{DAzstar}
D_{A}(z_{*})=\frac{c}{H_{0}}\frac{\mathrm{sinn}\left[H_{0}\sqrt{\left|\Omega_{K}\right|}\int_{0}^{z_{*}}\frac{dz}{H(z)}\right]}{(1+z)\sqrt{\left|\Omega_{K}\right|}} \, ,
\end{equation}
where $\mathrm{sinn(x)}=\sin(x)$, $x$, and $\sinh(x)$ for $\Omega_{K}<0$ ($k=1$), $\Omega_{K}=0$ ($k=0$), and $\Omega_{K}>0$ ($k=-1$), respectively. In the case of flat spatial current ($\Omega_{K}=0$), it is defined as
\begin{equation}
\label{DAzstar-flat}
D_{A}(z_{*})=\frac{c}{H_{0}(1+z)}\int_{0}^{z_{*}}\frac{dz}{\tilde{H}(z)} \, .
\end{equation}

In addition, $l_{a}$ is the angular distance of the sound horizon at the time of decoupling, and it is specified by the following equation
\begin{equation}
\label{la}
l_{a}=\left(1+z_{*}\right)\frac{\pi D_{A}(z_{*})}{r_{s}(z_{*})} \, .
\end{equation}
The coefficient $\left(1+z_{*}\right)$ appears because $D_{A}(z_{*})$ is the physical angular diameter distance in Eq. \eqref{DAzstar}, while $r_{s}(z_{*})$ is the associated comoving sound horizon at $z_{*}$ and is determined by Eq. \eqref{rs}.

Chen et al. \cite{Chen:2018dbv} have compared the analysis of the full CMB spectrum with the distance history method by constraining several dark energy models and have shown that the results of both methods are in perfect agreement. Therefore, in the present analysis, we consider the combined CMB likelihood (Planck 2018 TT, TE, EE + lowE) by considering $X_{i}^{\mathrm{obs}}=\{R,\,l_{a},\,\Omega_{b0}h^{2}\}=\{1.7502,\,301.4707,\,0.02236\}$, as determined by Chen et al. \cite{Chen:2018dbv}. The relation of $\chi_{\mathrm{CMB}}^{2}$ in this case is expressed as
\begin{equation}
\label{chi2CMB}
\chi_{\mathrm{CMB}}^{2}=\Delta X_{i}C_{ij}^{-1}\Delta X_{i}^{T} \, ,
\end{equation}
with $\Delta X_{i}=\left\{ X_{i}^{\mathrm{th}}-X_{i}^{\mathrm{obs}}\right\} $. The inverse of the covariance for this dataset is \cite{Chen:2018dbv}
\begin{equation}
\label{invcovCMB}
C_{ij}^{-1}=\left(\begin{array}{ccc}
94392.3971 & -1360.4913 & 1664517.2916\\
-1360.4913 & 161.4349 & 3671.6180\\
\,1664517.2916 & 3671.6180 & 79719182.5162
\end{array}\right) \, .
\end{equation}


\subsection{The SH0ES $H_0$ Measurement}
\label{subsec:H0}

Finally, we include the SH0ES team's \cite{Riess:2021jrx} measurement of the Hubble constant $H_0$, which quantifies the current expansion rate of the Universe. The Hubble constant is usually expressed as $H_{0}=100h\,\mathrm{km\ s^{-1}Mpc^{-1}}$. In the SH0ES project \cite{Riess:2021jrx}, the value of $H_0$ is determined by using observations of Cepheid variable stars and Type Ia supernovae as standard candles to establish precise distances to galaxies. By comparing these distances with the galaxies' redshifts, they derive a direct measurement of the Hubble constant. The inferred value from this project is \cite{Riess:2021jrx}
\begin{equation}
\label{H0_SH0ES}
H_{0}=73.04\pm1.04\,\mathrm{km\ s^{-1}Mpc^{-1}} \, .
\end{equation}
Then, the $\chi^2$ term is just
\begin{equation}
\label{chi2H0}
\chi_{H_{0}}^{2}=\left(\frac{H_{0}^{\mathrm{th}}-H_{0}^{\mathrm{obs}}}{\sigma_{H_{0}}}\right)^{2} \, ,
\end{equation}
where $H_{0}^{\mathrm{th}}$ indicates the result from the theoretical model.


\section{Results}
\label{sec:results}

To scan the parameter space of our model for the best fit to the observational data, we perform an MCMC analysis using the Metropolis-Hastings algorithm. Including the datasets described above, the total likelihood function in the MCMC analysis is
\begin{equation}
\label{Ltot}
\mathcal{L}_{\mathrm{tot}}=\mathcal{L}_{\mathrm{SN}}\times\mathcal{L}_{\mathrm{BAO}}\times\mathcal{L}_{\mathrm{CMB}}\times\mathcal{L}_{H_{0}} \, ,
\end{equation}
The total likelihood function is related to the quantity $\chi_{{\rm tot}}^{2}$ by $\chi_{{\rm tot}}^{2}=-2\ln\mathcal{L}_{\mathrm{tot}}$. Thus, we have
\begin{equation}
\label{chi2tot}
\chi_{{\rm tot}}^{2}=\chi_{{\rm SN}}^{2}+\chi_{{\rm BAO}}^{2}+\chi_{{\rm CMB}}^{2}+\chi_{H_{0}}^{2} \, .
\end{equation}

To examine the consistency of the standard cosmological model ($\Lambda$CDM) with the given datasets, the free parameters are $\{h,\Omega_{b},\Omega_{c}\}$. In our modified entropic gravity model, the free parameters are $\{h,\Omega_{b},\Omega_{c}\}$. Thus, our cosmological model has only one more degree of freedom than the $\Lambda$CDM model. The assumed priors for the free parameters of the $\Lambda$CDM model and our cosmological model are reported in Table \ref{table:priors}. The probability distributions for the starting values of each chain in all these priors are assumed to be uniform. With these priors, we performed our MCMC analysis and generated the Markov chains. For statistical analysis of the Markov chains, we used the publicly available GetDist package \cite{Lewis:2019xzd}.

\begin{table*}[!ht]
\centering
\caption{The priors for the free parameters used in our MCMC analysis. The priors of all the parameters are considered with a uniform probability distribution.}
\scalebox{1.0}{
\begin{tabular}{|l|c|c|}
\hline
\hline
Parameter $\qquad$ & $\qquad$ $\Lambda$CDM $\qquad$ & $\qquad$ MEC $\qquad$\tabularnewline
\hline
$h$ & $[0.6,0.8]$ & $[0.6,0.8]$\tabularnewline
$\text{\ensuremath{\Omega}}_{b0}$ & $[0.01,0.1]$ & $[0.01,0.1]$\tabularnewline
$\text{\ensuremath{\Omega}}_{c0}$ & $[0.1,0.5]$ & $[0.1,0.5]$\tabularnewline
$n$ & $-$ & $[1.0,20.0]$\tabularnewline
\hline
\end{tabular}
}
\label{table:priors}
\end{table*}

The results of our MCMC analysis for the best-fit values and 68\% CL constraints for the free parameters as well as the derived parameters in the studied models are listed in Table \ref{table:parameters}. In this table, we see that the modified entropic cosmology (MEC) model yields a Hubble constant value of $H_{0}=71.98\pm0.61\,\mathrm{km\ s^{-1}Mpc^{-1}}$, which is significantly larger than the result of the $\Lambda$CDM model, which yields $H_{0}=69.30\pm0.49\,\mathrm{km\ s^{-1}Mpc^{-1}}$. In contrast to the result of $\Lambda$CDM, the resulting value for this parameter in MEC overlaps with the 68\% CL constraint of SH0ES ($H_{0}=73.04\pm1.04\,\mathrm{km\ s^{-1}Mpc^{-1}}$) \cite{Riess:2021jrx}. Therefore, the MEC model can successfully resolve the Hubble tension. In Table \ref{table:parameters}, we see that the MEC model predicts lower values for the baryonic density and dark matter parameters. It is also clear from the table that our cosmological model yields a slightly lower age of the Universe compared to $\Lambda$CDM.

\begin{table*}[!ht]
\centering
\caption{The best-fit values and 68\% CL constraints for the parameters of the investigated models.}
\scalebox{1.0}{
\begin{tabular}{|l|c|c|c|c|}
\hline
\hline
\multirow{2}{*}{Parameter $\qquad$} & \multicolumn{2}{c|}{$\Lambda$CDM} & \multicolumn{2}{c|}{MEC}\tabularnewline
\cline{2-5}
 & best-fit & 68\% limits & best-fit & 68\% limits\tabularnewline
\hline
$h$ & $0.6932$ & $0.6930\pm0.0049$ & $0.7186$ & $0.7198\pm0.0061$\tabularnewline
$\text{\ensuremath{\Omega}}_{b0}$ & $0.04713$ & $0.04715\pm0.00055$ & $0.04322$ & $0.04308\pm0.00072$\tabularnewline
$\text{\ensuremath{\Omega}}_{c0}$ & $0.2474$ & $0.2477\pm0.0057$ & $0.2383$ & $0.2378\pm0.0054$\tabularnewline
$n$ & $-$ & $-$ & $6.55$ & $6.88_{-1.00}^{+0.64}$\tabularnewline
\hline
$H_{0}$ & $69.32$ & $69.30\pm0.49$ & $71.86$ & $71.98\pm0.61$\tabularnewline
$\text{\ensuremath{\Omega}}_{m0}$ & $0.2945$ & $0.2948\pm0.0062$ & $0.2815$ & $0.2809\pm0.0060$\tabularnewline
$\text{\ensuremath{\Omega}}_{\Lambda0}$ & $0.7054$ & $0.7051\pm0.0062$ & $-$ & $-$\tabularnewline
Age (Gyr) & $13.66$ & $13.666\pm0.022$ & $13.532$ & $13.530\pm0.024$\tabularnewline
$a_{c}$ & $-$ & $-$ & $0.8592$ & $0.8587\pm0.0095$\tabularnewline
$\tilde{\alpha}_{c}$ & $-$ & $-$ & $0.5186$ & $0.5181\pm0.0022$\tabularnewline
\hline
\end{tabular}
}
\label{table:parameters}
\end{table*}

In Table \ref{table:chi2}, MCMC results for the $\chi^2$ values for different datasets in the models studied are presented. This table shows that MEC fits with the CMB and SH0ES data better than $\Lambda$CDM. In turn, $\Lambda$CDM provides a better fit to the SN and BAO data. The value of $\chi_{\mathrm{total}}^{2}$ in MEC is lower than the corresponding value in $\Lambda$CDM, indicating that in general MEC fits the included data better.

In Table \ref{table:chi2}, we also report the values of the Akaike information criterion (AIC) \cite{Akaike:1974vps} for the models under study. This criterion is calculated using \cite{Akaike:1974vps}
\begin{equation}
\label{AIC}
\mathrm{AIC}=\chi_{{\rm tot}}^{2}+2k \, ,
\end{equation}
where $k$ represents the number of free parameters of the model under study. To evaluate the acceptability of these models, it is appropriate to assess the value of $\Delta\mathrm{AIC}=\mathrm{AIC}_{{\rm model}}-\mathrm{AIC}_{\Lambda\mathrm{CDM}}$. According to the interpretations for this criteria in \cite{BonillaRivera:2016use}: (1) if $\left|\Delta\mathrm{AIC}\right|\in(0,2]$, the considered model is substantially supported by the observational data, (2) if $\left|\Delta\mathrm{AIC}\right|\in[4,7]$, the observational data considerably less support the model, (3) if $\left|\Delta\mathrm{AIC}\right|>10$, the model is essentially not appropriate and should be discarded. As we see in Table \ref{table:chi2}, for the MEC model, $\left|\Delta\mathrm{AIC}\right|=0.67$ and therefore the AIC analysis states that this model is considerably supported by the observational data.

\begin{table*}[!ht]
\centering
\caption{The resulting values of $\chi^2$ for each model and each data set. The table also presents the results of AIC for the studied models.}
\scalebox{1.0}{
\begin{tabular}{|l|c|c|}
\hline
\hline
Parameter $\qquad$ & $\qquad$ $\Lambda$CDM $\qquad$ & $\qquad$ MEC $\qquad$\tabularnewline
\hline
$\chi_{{\rm SN}}^{2}$ & $1036.15$ & $1051.28$\tabularnewline
$\chi_{{\rm BAO}}^{2}$ & $2.78498$ & $4.81$\tabularnewline
$\chi_{{\rm CMB}}^{2}$ & $7.1604$ & $0.18$\tabularnewline
$\chi_{H_{0}}^{2}$ & $12.8033$ & $1.29$\tabularnewline
\hline
$\chi_{\mathrm{tot}}^{2}$ & $1058.90$ & $1057.57$\tabularnewline
AIC & $1064.9$ & $1065.57$\tabularnewline
\hline
$\Delta\chi^{2}$ & $0.0$ & $-1.33$\tabularnewline
$\Delta$AIC & $0.0$ & $0.67$\tabularnewline
\hline
\end{tabular}
}
\label{table:chi2}
\end{table*}

\begin{figure}
\centering
\includegraphics[width=\textwidth]{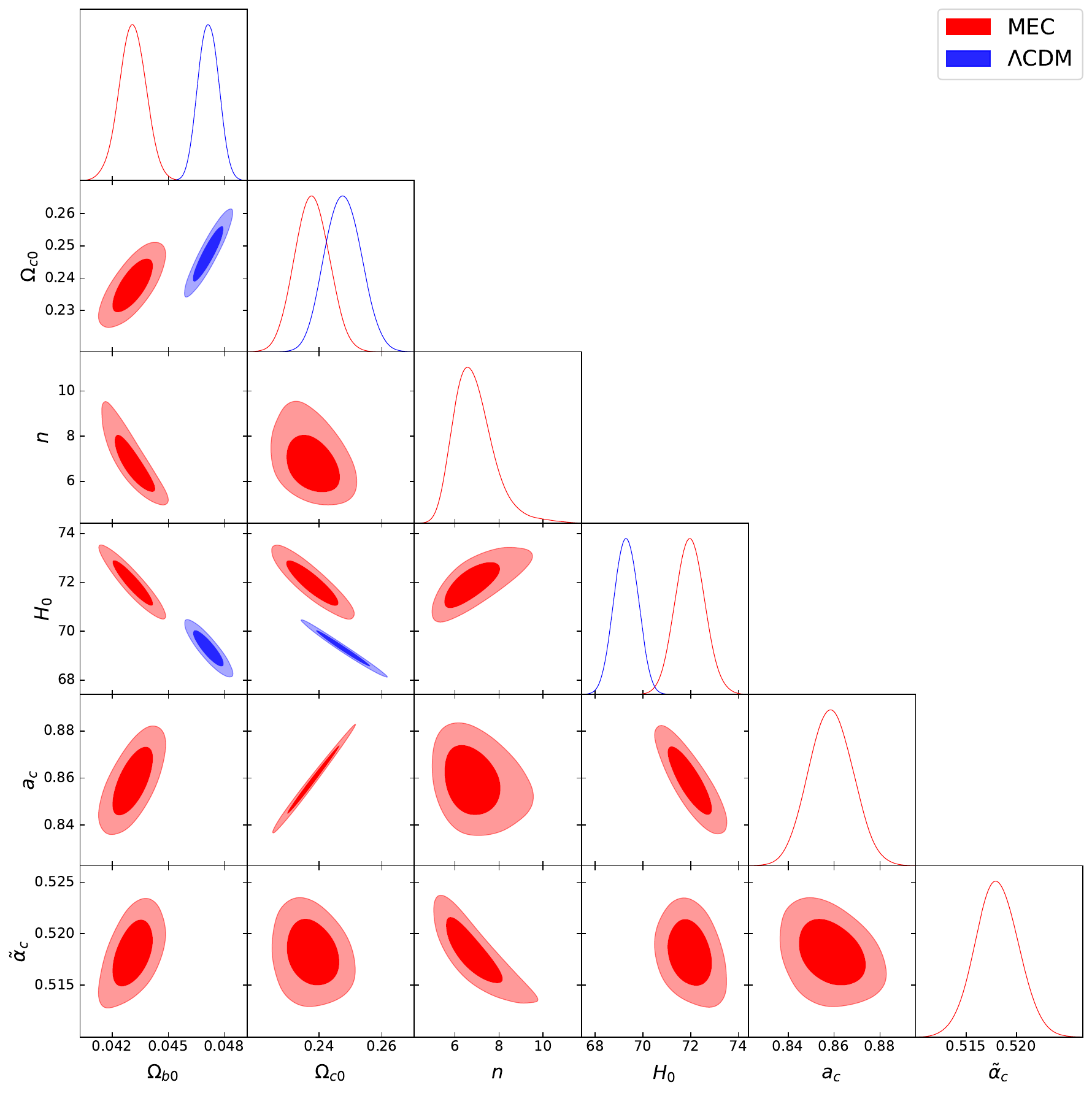}
\caption{1D likelihoods and 2D contours for the parameters in 68\% and 95\% CL marginalized joint regions for the $\Lambda$CDM model (blue) and the modified entropic cosmology (MEC) model (red).}
\label{fig:posteriors}
\end{figure}

The one-dimensional and two-dimensional posterior plots from our MCMC analysis are demonstrated in Fig. \ref{fig:posteriors}. It is also clear from this figure that MEC yields lower values for the baryonic energy density and cold dark matter compared to $\Lambda$CDM. It is also obvious from this figure that MEC yields a larger value for the parameter $H_0$ compared to $\Lambda$CDM, and hence has a better agreement with the SH0ES \cite{Riess:2021jrx} measurement. The figure also illustrates that the free parameter $n$, as well as the derived parameters $a_c$ and $\tilde{\alpha}_{c}$ can be well constrained by the included observational data so that their 68\% and 95\% CL contour-plot have been formed distinctively.

\begin{figure}
\centering
\includegraphics[width=0.6\textwidth]{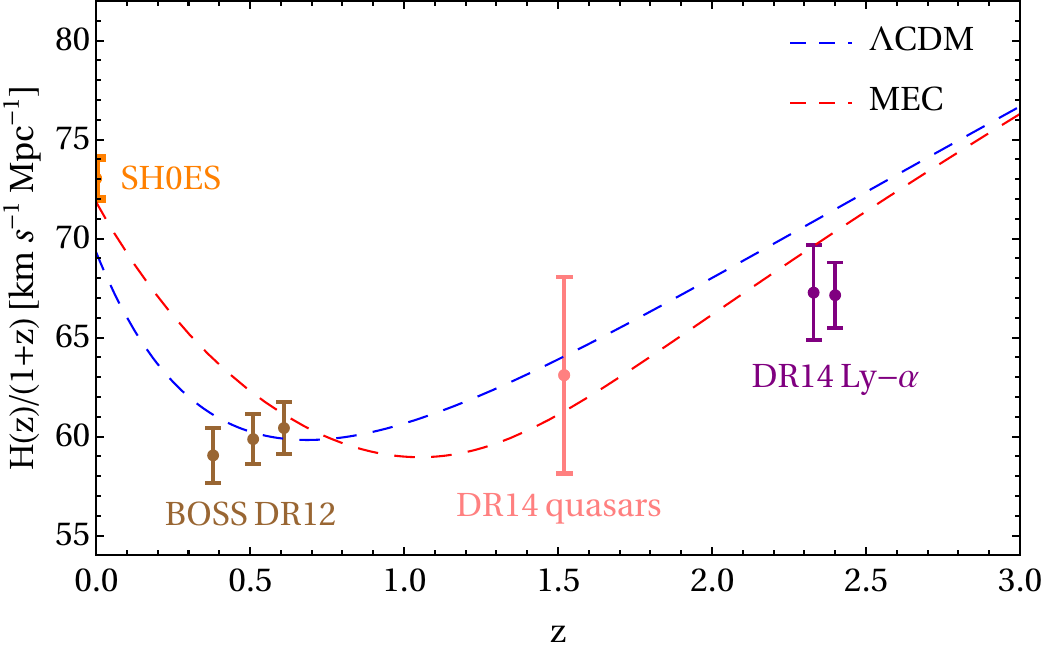}
\caption{Evolution of $H(z)/(1 + z)$ as a function of cosmological redshift in the $\Lambda$CDM and MEC models. In the figure, also the data points from SH0ES \cite{Riess:2021jrx}, BOSS DR12 \cite{BOSS:2016wmc}, DR14 quasars \cite{eBOSS:2018cab}, and DR14 Ly-$\alpha$ \cite{eBOSS:2019qwo} measurements have been specified for comparison.}
\label{fig:H}
\end{figure}

Figure \ref{fig:H} shows the variation of $H(z)/(1+z)$ with redshift in the setups of $\Lambda$CDM and MEC. To generate this plot, the best-fit values of the parameters in Table \ref{table:parameters} were used. This figure implies that at high redshifts, the diagram corresponding to MEC falls below the diagram of $\Lambda$CDM and therefore provides a better fit to the DR14 Ly-$\alpha$ data \cite{eBOSS:2019qwo}. At low redshifts, the plot of MEC lies above the plot of $\Lambda$CDM and therefore provides a better fit with the SH0ES measurement \cite{Riess:2021jrx}. However, at these redshifts, the plot of $\Lambda$CDM exhibits a better fit with the BOSS DR12 BAO measurements \cite{BOSS:2016wmc}.

\begin{figure}
\centering
\includegraphics[width=0.6\textwidth]{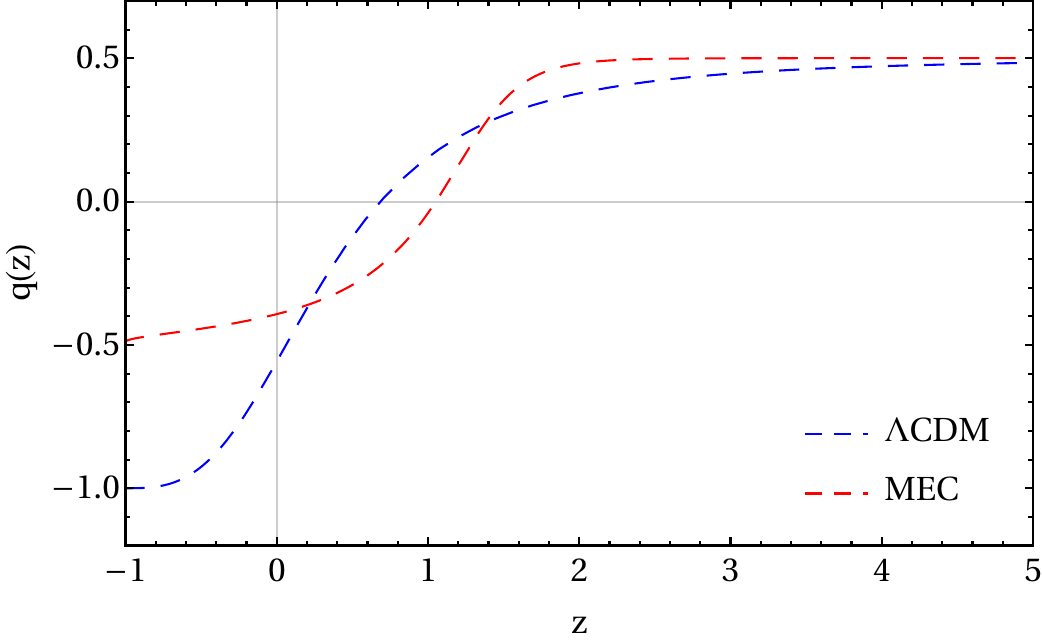}
\caption{Evolution of the deceleration parameter $q(z)$ as a function of cosmological redshift in the $\Lambda$CDM and MEC models.}
\label{fig:q}
\end{figure}

Figure \ref{fig:q} shows the variation of the deceleration parameter $q=-1-\dot{H}/H^{2}$ as a function of redshift for the two models studied. The main implication of this figure is that the MEC model can transit the accelerating phase near the present time. This result states that this model can explain the currently observed accelerating expansion of the Universe without any need for dark energy. Therefore, the cosmological constant problems that seriously challenge the standard model of cosmology do not arise at all in the case of our cosmological model. This means that MEC is free from the problems of the standard cosmological model, especially the fine-tuning problem, and this is a remarkable achievement for this scenario, which is based on the thermodynamical modifications of the gravity theory. Figure \ref{fig:q} also shows that MEC enters the accelerating phase earlier than $\Lambda$CDM. The MEC and $\Lambda$CDM models give the current value of the deceleration parameter as $-0.39$ and $-0.56$, respectively. In the figure, we see that the parameter $q$ goes towards the values $-0.49$ and $-1$ in the late future in the MEC and $\Lambda$CDM models, respectively.

\begin{figure}
\centering
\includegraphics[width=0.6\textwidth]{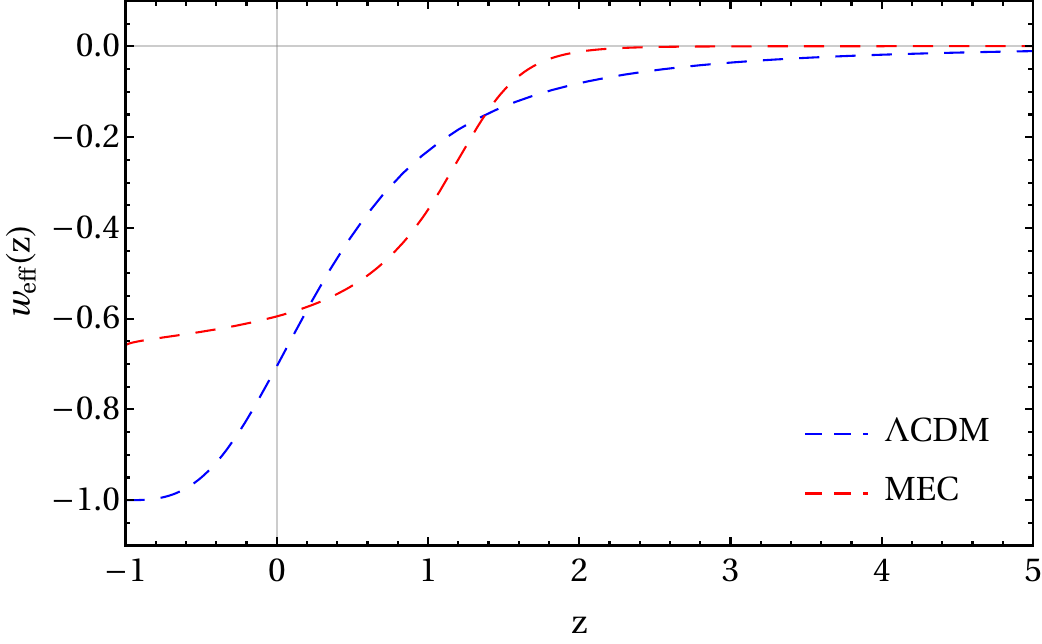}
\caption{Evolution of the effective equation of state parameter $w_{{\rm eff}}$ with redshift in the $\Lambda$CDM and MEC models.}
\label{fig:weff}
\end{figure}

In Fig. \ref{fig:weff}, the variation of the effective equation of state parameter $w_{{\rm eff}}=-1-2\dot{H}/3H^{2}$ is drawn versus redshift. In this figure, we see that the value of this parameter is currently equal to $-0.56$ and $-0.71$ for MEC and $\Lambda$CDM, respectively. Hence, the MEC model is currently in the quintessence regime ($w_{{\rm eff}} > -1$). In the late future, this model will also remain in the quintessence regime. The figure also illustrates that the final value of the parameter $w_{{\rm eff}}$ in the late future will be equal to $-0.66$ and $-1$ in the MEC and $\Lambda$CDM models, respectively. This implies that the MEC model behaves like a quintessence dark energy model. Recent data from DESI 2024 collaboration \cite{DESI:2024mwx} from the BAO measurements favor models with quintessence-like behavior, and so our model is supported by these observations as well.

At the end of this section, we discuss how the modified entropic cosmological model explains that the Universe should enter a phase of accelerated expansion at a time near the present. We know that in the periods of dominant radiation and dominant matter, the acceleration of the Universe was negative, and then at a time near the present the expansion of the Universe entered a phase of positive acceleration. So it is expected that at some time close to the end of the dominant matter period the acceleration would have become zero, and so the absolute value of the normalized acceleration would have become very small near that time. This means that near the time of changing the acceleration sign, the strength of the gravitational field would have been greatly reduced compared to earlier stages. It is expected that near the time of the acceleration sign change, there would have been a transition from a regime of strong gravitational fields to a regime of very weak gravitational fields. The MEC model requires that at this time the theory of gravity would have changed and would have diverged from Einstein's general relativity. So we observe that the accelerating phase of the Universe has occurred close to the present time because a very long time, about $13.5$ Gyr, has passed since the beginning of the Universe, and during this long period, the gravitational field strength has always been decreasing and has had enough time to reach very small values. As a result, the necessary conditions for the theory of gravity to change from Einstein's general relativity to modified entropic gravity have been met. In Table \ref{table:parameters}, we see that the value of the best fit of the critical normalized acceleration $\tilde{\alpha}_{c}$ is equal to $0.5186$. In this long time since the beginning of the Universe, the normalized acceleration has had enough time to decrease from very large values and reach this small value at a time before the present, and as a result, the conditions for changing the gravitational theory are fully met, and the gravitational theory that currently dominates requires that the Universe experienced a phase of positive acceleration with the current matter-energy content of the Universe. Thus, we see that within the framework of the MEC scenario, a convincing answer can be provided for the transition of the Universe to positive acceleration at a time close to the present.


\section{Conclusions}
\label{sec:conclusions}

In this work, we investigated the cosmological results of a model based on the modified entropic theory of gravity. This gravitational theory is proposed based on thermodynamic considerations of the gravitational force, and its idea is based on the concept of the entropic force proposed by Verlinde \cite{Verlinde:2010hp}. In our approach, we consider temperature corrections to the equipartition law of energy, which leads to the modification of Einstein's general relativity equations. We solved the modified Einstein equations for the homogeneous and isotropic FRW metric and derived the modified forms of the Friedmann equations. We confirmed, in a very interesting way, that the equations resulting from our general relativistic approach are consistent with those resulting from the thermodynamical corrections to the equations of classical Newtonian mechanics.

We applied the derived modified Friedmann equations to a flat Universe and investigated the evolution of the Universe. We found that the modified entropic cosmology (MEC) scenario can explain the current acceleration of the Universe without the need for any dark energy component. Since the modified entropic cosmological model is completely independent of dark energy, the problems related to the cosmological constant, especially the problem of fine-tuning, which seriously challenges the standard cosmological model, are generally not raised in the framework of this scenario.

We then constrained our cosmology model by using the observational data from different datasets including the Pantheon SN data \cite{Pan-STARRS1:2017jku}, BAO data \cite{Beutler:2011hx, SDSS:2009ocz, Blake:2011wn}, Planck 2018 CMB data \cite{Planck:2018vyg, Planck:2019nip, Planck:2018lbu}, and SH0ES measurement for $H_0$ \cite{Riess:2021jrx}. We found that our model can provide a better fit to the CMB and SH0ES data than the standard cosmological model, but the fit of the $\Lambda$CDM model to the SN and BAO data is better than our model. The MEC model results in $\chi_{\mathrm{tot}}^{2} = 1057.57$ which is smaller than the $\Lambda$CDM result which gives $\chi_{\mathrm{tot}}^{2} = 1058.90$ ($\Delta\chi^{2} = -1.33$). Therefore our model provides a better fit to the included observational data. The AIC analysis indicates that our model has a significant acceptance in light of the included data.

The MEC model yields a Hubble constant value of $H_{0}=71.98\pm0.61\,\mathrm{km\ s^{-1}Mpc^{-1}}$, which is significantly larger than the value of $H_{0}=69.30\pm0.49\,\mathrm{km\ s^{-1}Mpc^{-1}}$ that results from the standard cosmological model. This value is in agreement with the 68\% CL constraint from the SH0ES measurement \cite{Riess:2021jrx}, and thus, we see that the MEC scenario can successfully resolve the Hubble tension.

We saw that the MEC model enters the accelerating phase earlier than the $\Lambda$CDM model and the absolute value of the deceleration parameter $q$ is currently lower in our model than in the $\Lambda$CDM model. In the late future, the value of this parameter in the MEC and $\Lambda$CDM models is estimated as $-0.49$ and $-1$, respectively. The value of the effective equation of state parameter $w_{{\rm eff}}$ in the late future in the MEC and $\Lambda$CDM models is estimated to be $-0.66$ and $-1$ respectively. Therefore, the MEC model behaves like a quintessence-like dark energy, which is desirable according to the recent DESI BAO observations \cite{DESI:2024mwx}.

Within the framework of the modified entropic cosmological model, a convincing answer can be given for the transition of the Universe to a period of positive acceleration at a time near the present epoch. since the acceleration of the Universe has changed its sign near the present time, it is expected that the normalized acceleration at a time at the end of the matter domination era would have become very small. A very long time has passed since the beginning of the Universe, and the size of the Universe has expanded significantly. During this elapsed time, the acceleration has decreased intensively, which indicates that the strength of the gravitational field has decreased extremely. Thus, the conditions for the transition of the gravitational theory from Einstein's general relativity to the regime of very weak fields of modified entropic gravity are met. The regime of very weak fields of modified entropic gravity requires that the current expansion of the Universe be in the phase of positive acceleration with the current matter-energy content of the Universe.

In further developments of the current research, the cosmological perturbations theory can be investigated for the MEC model. For this purpose, it is necessary to perturb the modified Einstein equations. With this approach, the modified Boltzmann equations can be derived in this scenario and then the corresponding corrections can be applied in Boltzmann codes such as CAMB \cite{Lewis:1999bs} and CLASS \cite{Blas:2011rf}. In this case, we can calculate the angular power spectrum of CMB in this model and compare the results with the Planck 2018 observations \cite{Planck:2018vyg, Planck:2019nip, Planck:2018lbu} for the temperature and polarization anisotropies in the CMB radiation. Also, by investigating the perturbation theory for the late-time Universe, we can estimate the growth factor in this model and compare the results with the redshift-space distortion (RSD) surveys.

In addition, in further extensions of this research, other functions can be considered instead of the exponential function $f(x)$ that appears in the temperature correction relation of the energy equipartition law. As an important example, one can consider a function that behaves as $f\propto\tilde{\alpha}^{n}$,  in the limit of weak gravitational fields, where $n$ is a negative number. In this case, the final value of the effective equation of the state parameter of the model will be as $w_{{\rm eff}}=-2n/3(n+1)$ in the late times. This model can generate various scenarios for the dynamics of the late Universe. In this model, for $n < -3$, the model behaves like the quintessence dark energy model ($w_{{\rm eff}} > -1$) which is favored by the DESI BAO data \cite{DESI:2024mwx}. With $n=-3$, the model behaves like the cosmological constant ($w_{{\rm eff}} = -1$), and with $-3 < n < -1$, the model can provide a phantom-like behavior ($w_{{\rm eff}} < -1$). So, there are many possibilities within this scenario and therefore it may be possible to provide an even better fit to the observational data. This project is another interesting research that will be left to future investigations.











\bibliography{modified_entropic_cosmology} 


\end{document}